# A Software Framework for Vehicle-Infrastructure Cooperative Applications

Sebastian Bengochea, Angel Talamona, Michel Parent
INRIA, BP 105, 78153 Le Chesnay, France

*Abstract* - A growing category of vehicle-infrastructure cooperative (VIC) applications requires telematics software components distributed between an infrastructure-based management center and a number of vehicles. This article presents an approach based on a software framework, focusing on a Telematic Management System (TMS), a component suite aimed to run inside an infrastructure-based operations center, in some cases interacting with legacy systems like Advanced Traffic Management Systems or Vehicle Relationship Management. The TMS framework provides support for modular, flexible, prototyping and implementation of VIC applications. This work has received the support of the European Commission in the context of the projects REACT and CyberCars.

## I. INTRODUCTION

TELEMATICS, the technology that enables remote access to vehicle data over a mobile wireless network, regularly combines technologies of precise positioning using global navigation satellite systems and data communications using a cellular network. A wide category of vehicle-infrastructure cooperative (VIC) applications makes use of telematics software distributed between an infrastructure-based management center and the vehicles for safer and more efficient road transport solutions. The growing complexity and need of functionality integration of VIC systems require modular, reusable software components in order to improve application prototyping and implementation. This article presents an approach based on a software framework for VIC applications, focusing on a Telematic Management System (TMS), a component suite that usually runs inside an operations center, in some cases interacting with legacy systems, like Advanced Traffic Management or VRM systems as shown in Fig. 1.

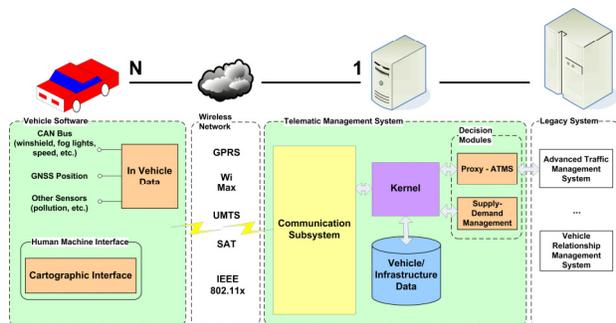

Fig. 1. The Telematic Management System is a link in the chain of vehicle-infrastructure applications. The TMS handles the connections of multiple vehicles and their data, providing services and integrating several applications (decision modules) over the same fleet.

The TMS framework provides infrastructure to integrate different decision modules, vehicle-management center communication support and an extensible data module. The pluggable decision modules are the key part of the TMS as they incarnate the logic for the VIC applications. The user of the framework develops and combines such modules for specific applications. Examples of decision modules include trip management for car-sharing, dynamic traffic modeling or proxy to third-party (eventually legacy) systems. A chain of dependencies is provided for each decision module and each module dependency is represented as a directed acyclic graph. The framework resolves, performing a topologic sort over the graph, the correct order of execution. An event system has been developed to maintain low coupling between the TMS framework and the decision modules interaction.

The framework provides the multithreading support for creating and handling the communication channels with the vehicles. The asynchronous communication is resolved using queues and a producer/consumer model between the different threads. Messages received from vehicles are propagated to the decision modules using an event system.

The data module provides a way to access both the vehicle (fleet/traffic) and infrastructure (cartography) data using interfaces. These modules can be extended by the user and provide a set of multi-threading tools for a synchronized data access.

The proposed architecture targets a current research need in vehicle-infrastructure cooperative systems [1].

Within the next section the TMS general architecture is presented. Section III contains the class and interface design for some of the key components of the architecture. Conclusion and further work follows.

## II. ARCHITECTURE OVERVIEW

This section presents a high level overview of the TMS framework architecture, describing briefly some of the design decisions taken for each part. The TMS is divided in three main modules; the Kernel, the Communication Subsystem (CS) and the Data Module. The user of the framework provides a set of decision modules and telematic protocol software to resolve the communication with the vehicles.

The Kernel defines a set of software interfaces that are used by the decision modules for accessing both the CS and the data module. This module also provides a set of mechanisms for handling concurrent access to the data. The Kernel initializes the server and the decision modules. To maintain low coupling between the TMS and the decision modules, an extensible event system has been developed. This low coupling approach allows dynamically adding and removing listeners without changing the predefined internal behavior of the TMS.

The decision modules and their dependencies are specified in an XML file that contains the id, the class that implements the decision module and their execution dependencies. This approach allows a declarative way for specifying the logic components used by the system. The execution dependencies are embedded as a list of decision modules that run before the current one is executed. This defines a direct acyclic graph of executing dependencies between all the decision modules. A topologic sort over this graph provides the execution order for each module (cyclic dependencies are not allowed in the current implementation). As previously stated, the modules are executed when an event (such as *vehicle logged in* or a *new message received* from vehicle) occurs, but it is also possible that the decision module runs in parallel to the TMS, for example working as proxy to a Trip Supply-Demand Management System for car-sharing. The decision modules are loaded in runtime, allowing the administrator of the system to replace or modify certain logic without stopping the execution of the VIC application.

The architecture and components that integrate a telematic system implemented using the framework is presented in Fig. 2.

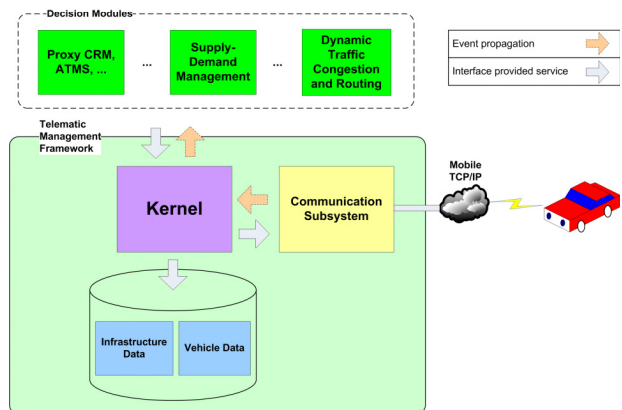

Fig. 2. The TMS framework propagate events to the Decision Modules, whereas these use the Kernel to access vehicle data or send messages to specific vehicles of the fleet.

The CS implements asynchronous communication for the TMS. The network protocol used is TCP/IP, which runs over a wide range of wireless links such as GPRS, Wi-Max, UMTS, SAT and IEEE 802.11x. Thus, the user of the framework defines the communication protocol used at the application level. Vehicles open a TCP/IP connection with the TMS and for each new connection established the CS creates a vehicle worker thread for handling the communication with the specific vehicle. For sending data to a vehicle, each vehicle worker contains a local queue and the worker sends data messages stored in this queue to the specific vehicle. The vehicle worker stores data received from the vehicle in a general queue that is part of the CS. The information of this queue is processed by certain number of dispatcher threads, each dispatcher pass the data message to the Telematic Protocol component.

The Telematic Protocol component is loaded at the startup of TMS. This component processes the incoming messages of the vehicles and may propagate events to the decision modules (or other listeners) each time a new message arrives. Different types of events can be defined and propagated by the user through this component. The user can provide different implementations of the Telematic Protocol (only one at a time is used). The latter allows different application-level protocols to operate with the TMS, such as SOAP-XML [2] or INRIA Lightweight Vehicular Telematics Protocol (LVTP) [3]. The CS internal composition is presented in Fig. 3.

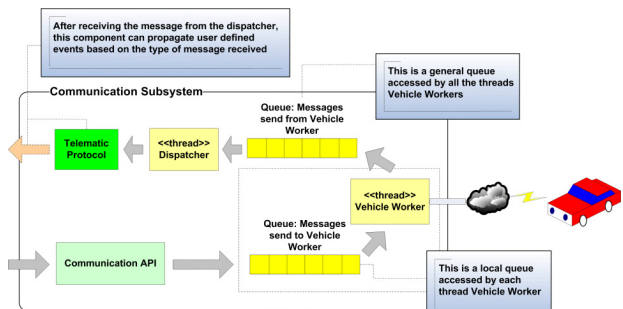

Fig. 3. The internal elements of the CS. The Communication API stores the received messages in the specific vehicle worker queue. When new messages arrive from the vehicle the worker stores this in a global queue. The dispatchers access the global queue, passing the message to the Telematic Protocol component, which is an event source propagating user specific events.

The proposed queue system, which is based on a producer-consumer model, makes possible to achieve certain level of asynchronous server-vehicle communication. It is also possible to distribute the work of the dispatchers and the vehicle workers under several servers, improving the scalability of the system.

## III. INTERFACE SPECIFICATION

This section will introduce the class and interface designs of the Event System, Decision Modules and the Telematic Protocol. The user of the framework provides some of these interfaces to implement different system behaviors. The TMS framework is currently implemented in Java.

### A. Event System

The event system used in the TMS is defined using three interfaces, the *ITMSEventSource*, *ITMSEventListener* and *ITMSEvent*, the UML class diagram for these interfaces is presented in Fig. 4.

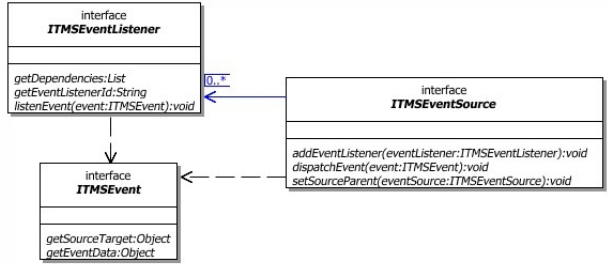

Fig. 4. Event system UML class diagram. The three central interfaces that define the event model system.

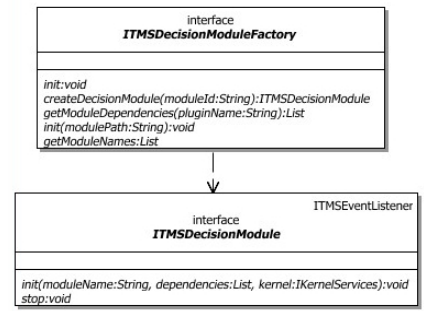

Fig. 5. UML class diagram for the decision modules. The decision module is initialized using the *init* method. Each instance will be registered by default to the events propagated through the TMS framework.

*ITMSEventListener* instances are registered to listen events from a specific event source (*ITMSEventSource* instances). Each *ITMSEventListener* instance provides the list of dependencies that should be executed before itself, defining a partial execution order for all the listeners.

*ITMSEventSource* instances are active components that propagate certain TMS or user defined events (*ITMSEvent* instances). Event source instances can be linked in a chain of event propagation, using the method *setSourceParent* of *ITMSEventSource*. When an event is propagated, the following protocol steps should be respected by each event source:

1. Executes default action event for the current event source.
2. Executes the *listenEvent* method for all registered listeners of the current event source, respecting the predefined order.
3. Propagates the event to the parent of the current event source (repeat the process from step 1).

*ITMSEvent* instances are the event objects that contain the event source target and some data related to the event. The event source target is the logic element of the TMS that generates the event, for example, if a *vehicle logged in* event is propagated then the event source target is the vehicle id.

### B. Decision Modules

The TMS framework instantiates decision modules using a factory configurable by an XML. A default implementation for the factory is provided by the TMS, this factory loads the XML information, resolves the dependencies and allows the runtime instantiation of each decision module. The UML class diagram for the decision modules is presented in Fig. 5.

Each declared decision module in the XML is an instance of *ITMSDecisionModule* interface. All the decision modules are initialized with the Kernel interface (API to the TMS framework), including the decision module name and dependencies list (information declared in the XML). The *ITMSDecisionModule* instance may run in parallel of the TMS (for example as a proxy to other legacy system), or be activated and executed only under certain events propagated.

### C. Telematic Protocol

The Telematic Protocol component processes the communication messages received from the vehicles, propagating the corresponding events in each case. This component is divided in three interfaces; *IMessage*, *IMessageFactory* and *ITelematicProtocol*.

The *IMessage* interface represents a message to exchange with the vehicle. The implementation of this interface provides the methods *marshal*/*unmarshal* which receive Input and Output Streams respectively. These methods allows the message to be serialized and deserialized using different formats provided by the user, making transparent to vehicle workers and dispatchers the real stream format exchanged over the network protocol. The *IMessageFactory* instance, provided as part of the Telematic Protocol component by the user, will be the creator of the *IMessage* instances. The *ITelematicProtocol* instance is responsible for processing the messages received from the vehicles, before different events may be propagated. A UML class diagram of the Telematic Protocol interfaces is presented in Fig. 6.

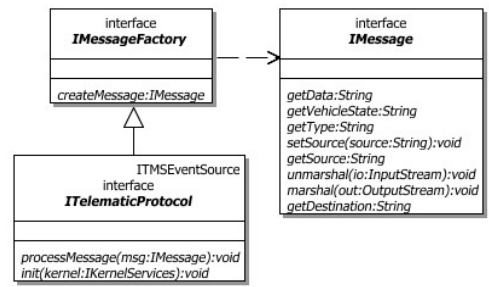

Fig. 6. Telematic Protocol UML class diagram. The user should provide an implementation of these interfaces, different implementation can support different telematic protocols.

The received messages are processed in two asynchronous process. The first process is handled by the vehicle worker that receives messages from the TCP connection. The received messages are stored in a general queue, after this the vehicle worker continues listening for new messages. This process is presented as an UML sequence diagram in Fig. 5.

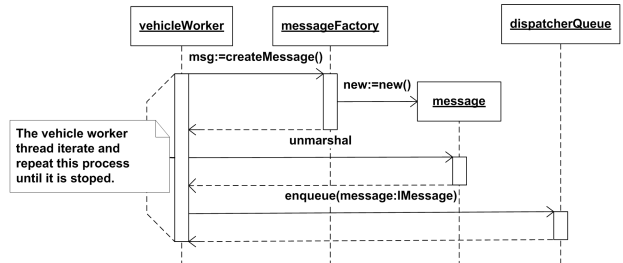

Fig. 7. Vehicle Worker UML Sequence Diagram. The *io* parameter used in the *unmarshal* method, is a stream of bytes received from the socket (previously opened by the communication subsystem). The *enqueue* method of the *IQueue* instance is non blocking. The iterative process is repeated until the vehicle worker thread is stopped.

The second process is handled by the dispatcher, if no messages are present in the general queue, then the dispatcher is blocked until a new message arrives. When a new message enters the general queue, it is passed by the dispatcher thread to the *ITelematicProtocol* instance for handling it. The dispatcher process is presented in the UML sequence diagram in Fig. 6.

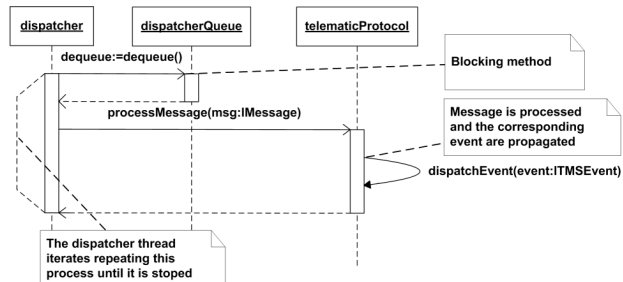

Fig. 8. Dispatcher UML sequence diagram. Each dispatcher thread iterates obtaining a message from the general queue. The message is passed to the *ITelematicProtocol* instance, which process the message and propagates user events.

Messages are sent to a specific vehicle (broadcast is also possible) creating and storing an instance of *IMessage* in the local queue of the specific vehicle worker. The vehicle worker will process the message through the TCP connection using the *marshal* output stream method.

IV. CONCLUSION AND FUTURE WORK

The TMS framework is currently under test at INRIA for in two different VIC applications. One of the applications manages a CTS (Cybernetic Transportation System) based on a fleet of automatic vehicles (cybercars). The fleet management system was tested with a fleet of three cybercars in the city of Nancy. The second one is an application with probe vehicles (driven cars equipped with sensors for atmospheric and infrastructure conditions). The cars transmit field data to the management system and get back safety warnings and route advises. Both applications share a dynamic traffic congestion model to send the vehicles route data in order to optimize the use of the road infrastructure, minimizing traffic congestion.

The current implementation of the queue system at the CS is a simple FIFO proposal while further versions of the TMS will implement distributed priority level queues for handling emergency or high priority messages. The TMS event system will be extended to work in a way similar to DOM Bubble event system [4]. The execution order of the decision modules, which is a direct acyclic graph (defined by their dependencies), will propagate events from decision modules to their fathers, and so on. This approach will improve the control of the decision modules over the propagated events, for example a module will be able to stop or modify an event under propagation.

Further versions of the TMS will work inside other standard servers, such as OSGi Server [5] or J2EE [6] application server. This will improve the framework with a set of advantages including component reusability, distributed deployment, integration with non-Java systems and application interoperability [7].

ACKNOWLEDGMENT

This work has received the support of the European Commission in the context of the projects CyberCars (www.cybercars.org) and REACT [8].